\documentclass{article}
\usepackage{fullpage}
\usepackage{graphicx}
\usepackage{geometry}
\usepackage{epstopdf}
\usepackage{amsmath}
\usepackage{amssymb}
\usepackage{listings}
\usepackage{color}
\usepackage{mwe}
\usepackage{hyperref}
\hypersetup{
	colorlinks=true,
	linkcolor=blue,
	filecolor=magenta,      
	urlcolor=cyan,
}

\definecolor{codegreen}{rgb}{0,0.6,0}
\definecolor{codegray}{rgb}{0.5,0.5,0.5}
\definecolor{codepurple}{rgb}{0.58,0,0.82}
\definecolor{backcolour}{rgb}{0.95,0.95,0.92}

\lstdefinestyle{mystyle}{
	backgroundcolor=\color{backcolour},   
	commentstyle=\color{codegreen},
	keywordstyle=\color{magenta},
	numberstyle=\tiny\color{codegray},
	stringstyle=\color{codepurple},
	basicstyle=\footnotesize,
	breakatwhitespace=false,         
	breaklines=true,                 
	captionpos=b,                    
	keepspaces=true,                 
	numbers=left,                    
	numbersep=5pt,                  
	showspaces=false,                
	showstringspaces=false,
	showtabs=false,                  
	tabsize=2
}

\author{Hardik Parwana\footnote{Research Assistant, IIT Kanpur email: parwanahardik@gmail.com}, Mangal Kothari\footnote{Assistant Professor, IIT Kanpur email: mangal@iitk.ac.in}\\ \textit{Department of Aerospace Engineering, Indian Institute of Technology Kanpur, India}}
\title{Quaternions and Attitude Representation}
\date{}
\begin{document}
	\maketitle

\section{Attitude Representation}

The attitude of a rigid body belongs to the configuration space known as Special Orthogonal group $ SO(3) $ and is represented in most general terms as $3\times 3$ rotation matrix. The space can, however, be parameterized in terms of fewer parameters. The advantages and disadvantages of some important representations used are summarized below:

\begin{itemize}
	\item \textbf{Rotation matrices}: This is the most general form for representing attitude of a body, also called special orthogonal group $ SO(3) $ group, is the space of $3\times 3$ matrices satisfying some constraints. Since it uses nine numbers to represent three angular degrees of freedom, there are six independent constraints on the matrix elements. Each column(and row) is unit vector, which gives us 3 constraints and the columns (and rows) are orthogonal to each other, yielding another 3 constraints. The translation ($\in R^3$) and rotation together are represented as Special Euclidean group $SE(3)$. The 6 constraints are larger compared to all other parameterizations mentioned below. They are therefore computationally more expensive than them. However, these have the advantage that they have no singularities or ambiguities such as double cover in attitude space in their representation as the rotation matrix is uniquely determined for a given configuration.
	
	\item \textbf{Euler angles}: These are more intuitive and easy to interpret physically. However, they suffer from singularity in mathematical formulations such as in Roll-Pitch-Yaw, also known as gimbal lock. At Pitch angle of $90^0$, it is unable to differentiate between Yaw and Roll degrees of freedom. There may also be a discontinuity when moving the angles in the parameter space, say, at $\pm\pi$. One of the Euler angles change suddenly in response to small change in rotation.
	
	\item \textbf{Unit Quaternions}: Owing to parametrization by 4 variables compared to 3 in Euler angles, these do not have singularity problem. Also, any rotation sequence can be represented by a continuous quaternion trajectory and do not suffer any discontinuity like Euler angles. These are also computationally more efficient than Euler angles since computation of sines and cosines is more expensive than simple operator on numbers. However, they possess the ambiguity of dual covering, i.e., the quaternions $q$ and $-q$ represent the same quaternion. It is also easier to interpolate between rotations and to chain rigid transformations.
		
	\item \textbf{Axis/angle representation}: It parameterizes the rotation by a unit vector $\vec
	n$ and a rotation about it by angle $\theta$. Since unit vector has a norm constraint, three parameters are required for representation, 2 for unit vector and 1 for rotation angle. The representation is not unique and an additional rotation of $360^0$ gives the same rotation matrix. However it has many advantages too. The representation is minimal and does not require any constraints on parameters such as unit modulus of quaternions used to represent rotations. It is easer to understand the pose (example a twist about y axis) and the derivatives of rotation matrix R with respect to rotation angle can easily be computed.\\
\end{itemize}

\section{Problem with Euler Angles}
Although Euler angles are widely used because of their minimalistic (3 parameters for 3 Degrees of Freedom) and intuitive physical representation, they exhibit a phenomenon known as \textit{Gimbal Lock} which results in a singularity in the attitude representation. Note that the Euler angles are relative sequential rotations (See video on GuerrilaCG Youtube channel: \href{https://www.youtube.com/watch?v=zc8b2Jo7mno
	}{Euler (gimbal lock) Explained: https://www.youtube.com/watch?v=zc8b2Jo7mno
	}). For example in a 123 rotation, a single roll angle will change the orientation of pitch and yaw axis  but in case of a single pitch angle rotation, the roll axis remains intact while yaw axis changes its orientation with the pitch rotation. When the pitch is $90^0$, the roll and yaw axis becomes the same and system can rotate only about 2 axis in space at that particular instant. Therefore, when the axis of two out of the three gimbals are driven parallel to each other in a configuration, there is a loss of degree of freedom and the dimension of the attitude space reduces to 2. The gimbal is still free to move(there is no actual mechanical \textit{lock} as the name suggests) however, in order to move along the third missing axis in this configuration, the body will have to move simultaneously along two gimbals axis and may exhibit unfamiliar motions when such a situation is encountered physically. There are many ways in which people deal with this problem while still using Euler angles however, that treatment is not a topic of these notes.

\subsection{Mathematical Implication}
Mathematically the formulation of attitude space in terms of Euler angles can cause a singularity and demand .... commands. Consider a rigid body with its angular velocity represented by the orthogonal vector $[p~q~r]^T$ as per the convention. The Euler rates $[\dot{\phi} ~ \dot{\theta} ~ \dot{\psi}]^T$ are not orthogonal as they are defined in different frame of references and, for a 321 rotation, can be related to angular velocity as follows:

\begin{eqnarray}
		\begin{bmatrix}
		p\\q\\r \end{bmatrix} = \begin{bmatrix}
		\dot{\phi}\\0\\0
		\end{bmatrix}+ R(\phi)\begin{bmatrix}
		0\\\dot{\theta}\\0
		\end{bmatrix} + R(\phi)R(\theta)\begin{bmatrix}
		0\\0\\\dot{\psi}
		\end{bmatrix}\\
		\begin{bmatrix}
		p\\q\\r \end{bmatrix}  =  \begin{bmatrix}
		1 & 0 & -\sin\theta\\
		0& \cos\phi & \sin\phi\cos\theta\\
		0 & -\sin\phi & \cos\phi\cos\theta 
		\end{bmatrix}\begin{bmatrix}
		\dot{\phi}\\\dot{\theta} \\ \dot{\psi}
		\end{bmatrix}
\end{eqnarray}
The determinant of the above matrix becomes zero at $\theta=90^0$ and there the dimension of the space reduces to 2. The matrix can be inverted to obtain Euler rates in terms of angular velocity as:
\begin{equation}
		\begin{bmatrix}
		\dot{\phi}\\\dot{\theta} \\ \dot{\psi}
		\end{bmatrix} = \begin{bmatrix}
		1 & \sin\phi\tan\theta & \cos\phi\tan\theta \\
		0 & \cos\phi & -\sin\phi \\
		0 & \sin\phi\sec\theta & \cos\phi\sec\theta
		\end{bmatrix}\begin{bmatrix}
		p\\q\\r \end{bmatrix}
\end{equation}
At the singularity point, the matrix elements are not defined as $\tan$ and $\sec$ terms go to $\infty$.

\section{Quaternions}

Quaternions can be considered as 4 component extended complex number of the form 
\begin{equation}
\textbf{q} = q_0 + q_1i + q_2j + q_3k  ~~~~  p\circ q \neq q\circ p ~~~~     p\circ q\circ r = (p\circ q)\circ r = p\circ (q\circ r)
\end{equation}
where in $q=(q_0,\vec{q})$, $q_0$ is called the real part and $\vec{q}=(q_1~q_2~q_3)^T$ is called the imaginary part of the quaternion. Some concepts and definitions used in Quaternion theory are described below[1]:

	\begin{itemize}
		\item A right handed notation follows the following rules:
		\begin{eqnarray}		
		i^2 &=& j^2 = k^2 = ijk = -1\\
		ij &=& k, ji = -k\\
		jk &=& i, kj = -i\\
		ki &=& j, ik = -j
		\end{eqnarray}
		
		\item The conjugate $\bar{q}$ of $q$ is defined as 
		\begin{equation}
		\bar{q} = (q_0,-\vec{q})
		\end{equation}
		
		\item Norm is defined as
		\begin{equation}
		|q| = \sqrt{q_0^2 + q_1^2 + q_2^2 + q_3^2}
		\end{equation}
		
		\item The quaternion product is denoted as $q\circ p$ and can be obtained using the rules of Eqs. 2-5.
		\begin{eqnarray}
		q\circ p &=& (p_0q_0-\vec{p}\cdot\vec{q},q_0\vec{p}+p_0\vec{q}+\vec{q}\times\vec{p})\\
		(q_0 + q_1i + q_2j + q_3k)(p_0 + p_1i + p_2j + p_3k)
		&=&\begin{pmatrix}
		p_0q_0 - q_1p_1 - q_2p_2 - q_3p_3\\
		q_1p_0 + q_0p_1 + q_2p_3 - q_3p_2\\
		q_2p_0 + q_0p_2 + q_3p_1 - q_1p_3\\
		q_3p_0 + q_0p_3 + q_1p_2 - q_2p_1
		\end{pmatrix}\\			
		&=& \begin{bmatrix}
		q_0 &-q_1 &-q_2& -q_3\\
		q_1 &q_0 &-q_3& q_2\\
		q_2 &q_3& q_0 &-q_1\\
		q_3 &-q_2& q_1 &q_0\\
		\end{bmatrix}\begin{pmatrix}
		p_o\\p_1\\p_2\\p_3
		\end{pmatrix}\\
		&=& Q(q)p\\			
		&=& \begin{bmatrix}
		p_0 &-p_1& -p_2& -p_3\\
		p_1 &p_0& p_3 &-p_2\\
		p_2 &-p_3& p_0 &p_1\\
		-p_3& p_2 &-p_1 &p_0
		\end{bmatrix}\begin{pmatrix}
		q_0\\q_1\\q_2\\q_3
		\end{pmatrix}\\			
		&=& \bar{Q}(p)q
		\end{eqnarray}
		
		\item \begin{eqnarray}
		\bar{q\circ p} &=& \bar{p}\circ \bar{q}\\
		|q \circ p| &=& |q||p|
		\end{eqnarray}
		
		\item Inverse of a quaternion is defined as :
		\begin{equation}
		q^{-1} = \frac{\bar{q}}{|q|}
		\end{equation}
		This is similar to inverse in complex numbers:
		\begin{equation}
		(a+ib)^{-1} = \frac{a-ib}{\sqrt{(a^2+b^2)}}
		\end{equation} 
		
	\end{itemize}
	
	\section{Quaternion: Intuitive Representation}
	
	A unit quaternion ($|q|=1$) can be used to represent a rotation in space. This set of 4 dimensional vectors with unit modulus are said to belong to set $S^3$[4]. Since its modulus is one, $q^{-1}=\bar{q}=q^*$($q^*$ is also the notation for conjugate of quaternion which is more commonly used in case quaternion elements are complex numbers. Since rotation quaternions have real entries, $q^*=\bar{q}$). Henceforth, all the quaternions mentioned will be unit quaternions unless stated otherwise. \\ \\	
		\textbf{Euler's Rotation Theorem}\\
		\textbf{Any rotation or sequence of rotations of a rigid body or coordinate system about a fixed point is equivalent to a single rotation by a given angle θ about a fixed axis (called Euler axis) that runs through the fixed point.}
		
		Quaternions can be very easily correlated to the axis angle representation of attitude. Any attitude of a rigid body can be defined by stating an axis in 3D with unit vector $\vec{n}$, and a rotation about that axis, $\theta$. A rotation quaternion can be represented in terms of these two as:
		\begin{equation}
			q =  \left(cos\left(\frac{\theta}{2}\right),\vec{n}\sin\left(\frac{\theta}{2}\right)\right)
		\end{equation}
		
		As verification, consider the well known rotation formula studied in dynamics course.
		\begin{equation}
			\vec{x}' = (1-\cos(\theta))(\vec{n}.\vec{x})\vec{n} + \cos\theta\vec{x} + \sin\theta(\vec{x}\times\vec{n})
		\end{equation}
		where $\theta$ is the rotation about the axis of rotation $\vec{n}$
    
    Now consider the equation
    \begin{equation}
    	x' = \bar{q}\circ x\circ q ~~~~~~ x=(0,\vec{x}) ~~~~~~~ q = (q_0,\vec{q})
    \end{equation}    	
    
    Simplifying the above, we get
    	\begin{eqnarray}
    	Re(x') &= &(\vec{q}.\vec{x})q_0 - q_0(\vec{x}.\vec{q}) - (\vec{q}\times\vec{x})\cdot q = 0\\
    	Im(x') &=& 2(\vec{q}.\vec{x})\vec{q} + q_0^2\vec{x} + 2q_0(\vec{x}\times\vec{q}) - (\vec{q}.\vec{q})\vec{x}
    	\end{eqnarray}
    	
   	If we now choose $ q $ as $q=(\cos\frac{\theta}{2},\sin\frac{\theta}{2}\vec{n})$, $|n|=1$, then $ x' $ becomes:
    	
    	\begin{equation}
    	\vec{x'} = (1-\cos(\theta))(\vec{n}.\vec{x})\vec{n} + \cos\theta\vec{x} + \sin\theta(\vec{x}\times\vec{n})
    	\end{equation}    	
   	This is same as the known rotation formula.
   	
   	\subsection{Extension of complex numbers}
\begin{figure}
   			\centering
   			\includegraphics[width=0.9\textwidth]{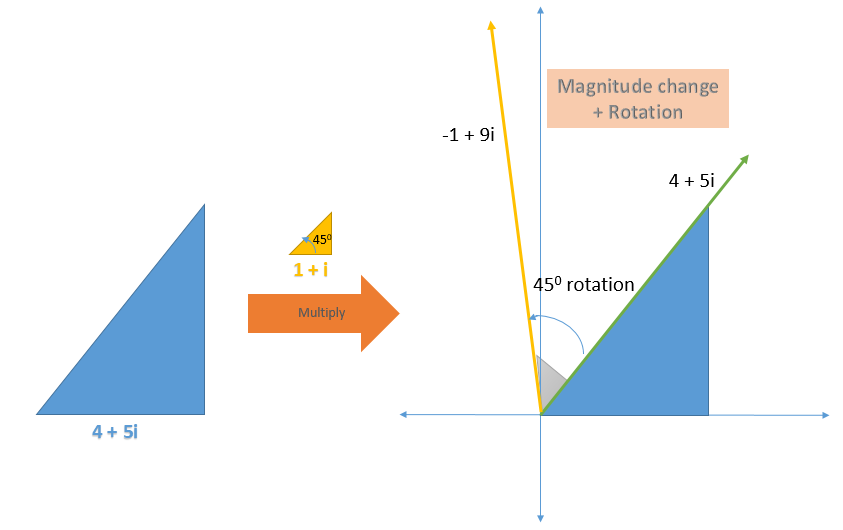}
   			\caption{Complex number multiplication}
   			\label{eg:complex}
   		\end{figure}

   	It is well known that multiplying a complex numbers $r_1e^{i\theta_1}$(=$r_1\cos\theta_1 + ir\sin\theta_1$) by $r_2e^{i\theta_2}$ results in amplification of its magnitude by $r_2$ times and change in orientation by an angle $\theta_2$ (product = $r_1r_2e^{i(\theta_1+\theta_2)}$). See Fig.\ref{eg:complex} for an example. Now if we were to multiply with a unit modulus complex number, i.e. $r_2=1$, then we can perform pure rotation. Quaternions can be seen just to be an extension of this mathematical construction and the constraint of unit modulus to higher dimension.
   	
	\section{Quaternion Conventions}
	Quaternions are represented in many ways and have different interpretations. For example, one can consider rotating frames or rotating vectors or a left handed or right handed operations(Note that the usual vector cross product is based on right handed system, however it can also be defined in a left handed system). Two most commonly used notations are Hamilton and JPL conventions, both of which considers rotating frame[2]. The one used in these notes above and which will be used in subsequent theory is the Hamilton convention. Some differences between two are:\\ \\
	
	\begin{table}
		\centering
		\begin{tabular}{c|c|c}
			Quaternion Type & Hamilton & JPL \\
			\hline
			1. Component order & $ (q_0,\vec{q}) $ & $ (\vec{q},q_0) $\\
			2. Algebra & ij=k(right handed) & ij=-k(left handed)\\
			3. Default notation & Local to Global & Global to local \\
			  & $ q = q_{GL} $ & $ q=q_{LG} $ \\
			  & $ x_G = q\circ x_L\circ q* $ &$  x_L = q\circ x_G \circ q* $\\
			  \hline
		\end{tabular}
	\end{table}
	where $L$ stands for local frame and $G$ stands for global frame. Therefore given the sequence of operation $q\circ - \circ q^*$, we go from local frame to global frame in Hamilton notation and vice versa in JPL when going from left hand side to right hand side in an equation. 
	
	\section{Quaternion Derivatives}
	
	Quaternion derivatives are often used to represent angular velocity or to integrate the system dynamics. This section will state and derive some important relations which are very common in literature.
	
	\subsection{First Derivative}
	
	The quaternion rates are given in terms of angular velocity expressed in global frame ($w = (0,\vec{w})$) or body frame ($w' = (0,\vec{w}')$).
	\begin{eqnarray}
	\dot{q} &=& \frac{1}{2}q\circ w'\\
	\dot{q} &=& \frac{1}{2}w \circ q\\
	w' &=& 2\bar{q}\circ \dot{q}\\
	w &=& 2\dot{q}\circ \bar{q}
	\end{eqnarray}
	
	A simple derivation for the quaternion rate is done using the first principle[2].
	
	\begin{eqnarray}
	\dot{q} &=& \lim\limits_{\triangle t\rightarrow 0}\frac{q(t+\triangle t)-q(t)}{\triangle t} \\
	& = & \lim\limits_{\triangle t\rightarrow 0}\frac{q\circ \triangle q_L-q(t)}{\triangle t} \\
	& = & \lim\limits_{\triangle t\rightarrow 0}\frac{q\circ (\begin{bmatrix}
		1\\ \vec{n}\triangle\theta_L/2
		\end{bmatrix}-\begin{bmatrix}
		1\\0
		\end{bmatrix})}{\triangle t} \\
	& = & \lim\limits_{\triangle t\rightarrow 0}\frac{q\circ (\begin{bmatrix}
		0\\ \vec{n}\triangle \theta_L/2
		\end{bmatrix})}{\triangle t} \\
	& = & \frac{1}{2}q\circ \begin{bmatrix}
	0 \\ \vec{w}' \label{qder}
	\end{bmatrix}
	\end{eqnarray}
	
	Note that in Hamilton notation, local disturbances or rotations(i.e., with respect to current configuration) are post-multiplied as done above since we move from local to global frames when going left. Thus, the change in angle $\triangle\theta_l/\triangle t$ represents the angular velocity in body frame. Similarly if one were to add global disturbances(i.e., with respect to glbal coordinate system and not the current body configuration), they would be pre-multiplied and not post-multiplied. \textbf{This can be better understood by following example. } Consider the following two sets of consecutive rotations:
	\begin{enumerate}
		\item \textbf{Case a:} Rotation about $ Z $ by $45^0$, then rotation about $ X' $(new $ X $ axis) by $90^0$
		\item \textbf{Case b:} Rotation about $ Z $ by $45^0$. then rotation about $ X $(original $ X $ axis) by $90^0$
	\end{enumerate}
	
\begin{figure}
		\centering
		\includegraphics[width=0.9\textwidth]{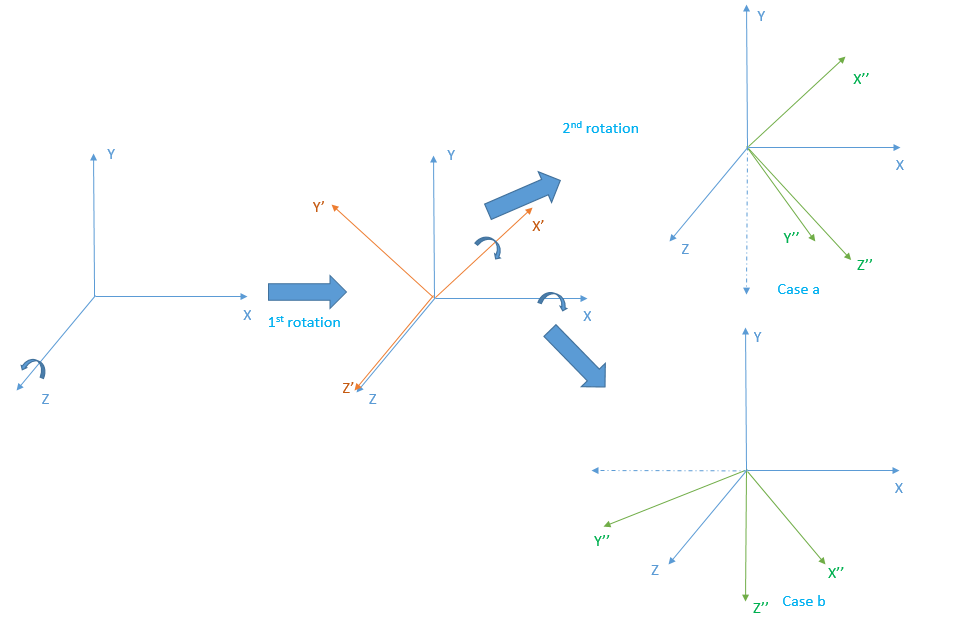}
		\caption{Quaternion local and global rotation perturbation example}
		\label{eg}
	\end{figure}

	Consider the vector $[0~0~1]$ in the final frame of reference. We desire to obtain this vector in the original frame of reference. From Fig.\ref{eg} it is clear that \textbf{Case a} should yield $[\frac{1}{\sqrt{2}} ~ \frac{-1}{\sqrt{2}} ~ 0]$ as the answer and \textbf{Case b} should yield $[0~-1~0]$ as the answer.
	
	Let us denote first rotation quaternion by $q^1$ and the second by $q^2$. In the first case, since we are rotating with respect to local frame of reference, this local perturbation is post multiplied, i.e., $q^a=q^1\circ q^2$. Whereas, in the second case, we are rotating again with respect to global frame of reference, therefore we will be pre-multiplying, i.e., $q^b=q^2\circ q^1$.
	
\begin{lstlisting}[language=MATLAB, caption=MATLAB code]
q1 = [cos(pi/8);0;0;sin(pi/8)]; %rotation about Z axis by 45 degrees
q2 = [cos(pi/4);sin(pi/4);0;0]; %rotation about X axis by 90 degrees

qa = quatmultiply(q1,q2)
qb = quatmultiply(q2,q1)

x = [0;0;1]; %vector in final frame of reference

%vector in global frame of reference for Case a
xa = quatmultiply(qa,quatmultiply([0;x],[qa(1);-qa(2);-qa(3);-qa(4)]))

%vector in global frame of reference for Case b
xb = quatmultiply(qb,quatmultiply([0;x],[qb(1);-qb(2);-qb(3);-qb(4)]))

function [ Q ] = quatmultiply( q,p )
%Performs Quaternion multiplication q o p
        q0=q(1);
        q1=q(2);
        q2=q(3);
        q3=q(4);

        Q1 = [q0 -q1 -q2 -q3;
        q1  q0 -q3  q2;
        q2  q3  q0 -q1;
        q3 -q2  q1  q0]; 
        Q = Q1*p;
end		
\end{lstlisting}
		
	\begin{lstlisting}[language=MATLAB, caption=Matlab Code Result]
		qa =
		
		0.6533
		0.6533
		0.2706
		0.2706
		
		
		qb =
		
		0.6533
		0.6533
		-0.2706
		0.2706
		
		
		xa =
		
		0
		0.7071
		-0.7071
		0.0000
		
		
		xb =
		
		0
		0
		-1.0000
		0.0000
	\end{lstlisting}

	Let $x$ denote the quaternion corresponding to position vector in global frame and $x'$ denote the quaternion corresponding to position vector in local frame. Another derivation for the derivative formula is as follows[1]:
	 \begin{eqnarray}
	x &=& q \circ x' \circ \bar{q}\\
	\dot{x} &=& \dot{q}\circ \bar{q} \circ x \circ q \circ \bar{q} + q\circ \bar{q} \circ x \circ q \circ \dot{\bar{q}}\\
	\dot{x} &=& \dot{q}\circ x' \circ \bar{q} + q\circ x'\circ \dot{\bar{x}}\\
	&=&\dot{q}\circ\bar{q}\circ x + x\circ q\circ \dot{\bar{q}}
	\end{eqnarray}
	Now from definition of quaternion product,
	\begin{eqnarray}
		\dot{q} \circ \bar{q} &=& (\dot{q}_0q_0+\dot{\vec{q}}\cdot\vec{q},-\dot{q}_0\vec{q}+q_0\dot{\vec{q}}-\dot{\vec{q}}\times\vec{q})\\
		&=& (0,\vec{v})
	\end{eqnarray}
	where in last step scalar part becomes zero, which can be obtained by differentiating $|q|^2=1$, and the vector part is defined as $\vec{v}$ for convenience.
	
	Therefore, we get
	\begin{eqnarray}
		\dot{x} &=& (0,\vec{v})\circ x - x\circ (0,-\vec{v})\\
		&=&2\vec{v}\times \vec{x}
	\end{eqnarray}
	Note that it can be easily shown from above expressions and using $|p\circ q|=|p||q|$ that $|\dot{\vec{x}}|=|2\vec{v}||\vec{x}|$ which implies that $\vec{v}$ is perpendicular to $\vec{x}$.
	
	Now for pure rotation, $\dot{\vec{x}} = \vec{w}\times \vec{x}$ where $\vec{w}$ is the angular velocity and is perpendicular to $\vec{x}$. Therefore, we have	
	\begin{eqnarray}
	\vec{w} &=& 2\vec{v}\\
	\implies w &=& (0,\vec{w}) = (0,2\vec{v}) = 2\dot{q}\circ \bar{q}\\
	\implies \dot{q} &=& \frac{1}{2}w\circ q
	\label{qdotww}
	\end{eqnarray}
	
	Now the angular velocity in body frame is given by $w' = \bar{q}\circ w\circ q$ and hence Eq.(\ref{qdotww}) becomes
	\begin{equation}
		\dot{q} = \frac{1}{2}q\circ w'
	\end{equation}

	It is very convenient to have matrix notations for all the operations mentioned above. Let $G$ and $E$ be defined by:
	\begin{eqnarray}
	E &=& \begin{pmatrix}
	-q_1& q_0& -q_3& q_2\\
	-q_2 &q_3 &q_0 &-q_1\\
	-q_3 &-q_2 &q_1& q_0
	\end{pmatrix}\\
	G &=& \begin{pmatrix}
	-q_1& q_0& q_3& -q_2\\
	-q_2 &-q_3 &q_0 &q_1\\
	-q_3 &q_2& -q_1& q_0
	\end{pmatrix}
	\end{eqnarray}
	
	The angular velocities can then also be written in terms of G and E as follows:
	\begin{eqnarray}
	w &=& 2E\dot{q}\\
	&=&-2\dot{E}q\\
	w' &=& 2G\dot{q}\\
	&=& -2\dot{G}q
	\end{eqnarray}
	
	It can easily be shown that following also holds:
	\begin{eqnarray}
		\dot{q} = \frac{1}{2}E^T\vec{w} = \frac{1}{2}G^T\vec{w}'
	\end{eqnarray}
	
	Note that $\vec{w}=2E\dot{q}=2E(E^T\vec{w}/2)=EE^T\vec{w}$. Therefore $EE^T=I$. Similarly $GG^T=I$.	
	
	\subsection{Second Derivative}
	
	We will now derive relationship between quaternion and angular acceleration. Differentiating $w'=2\bar{q}\circ \dot{q}$, we get
	\begin{eqnarray}
		\dot{w}' &=& 2\bar{q}\circ\ddot{q} + 2\dot{\bar{q}}\circ\dot{q}\\
		 &=& 2\bar{q}\circ\ddot{q} + 2\begin{bmatrix}
		 ||\dot{q}||^2\\0\\0\\0
		 \end{bmatrix}\\
		 &=& 2Q(\bar{q})\ddot{q} + 2\begin{bmatrix}
		 ||\dot{q}||^2\\0\\0\\0
		 \end{bmatrix}\\
		 \implies \dot{\vec{w}}' &=& 2G\ddot{q}\\
		 where~ G^T &=& \begin{bmatrix}
		   -q_1 & -q_2 & -q_3\\
		   q_0 & -q_3 & q_2 \\
		   q_3 & q_0 & -q_1\\
		   -q_2 & q_1 & q_0
		   \end{bmatrix}
	\end{eqnarray}
	
	One can similarly work out the derivative in terms of world frame angular velocity $w$.
	
	\section{Rotation Matrix}
	
		The rotation matrix R which transforms $\vec{x}'$ to $\vec{x}$ can be written as[1]:
		\begin{equation}
		R = EG^T		
		\end{equation}
		
		The R matrix is obtained as:
		\begin{equation}
		R = \begin{bmatrix}
		q_0^2+q_1^2-q_2^2-q_3^2 &   2(q_1q_2-q_0q_3)   &      2(q_0q_2+q_1q_3)\\
		2*(q_1q_2+q_0q_3)&   (q_0^2-q_1^2+q_2^2-q_3^2) &   2(q_2q_3-q_0q_1)\\
		2(q_1q_3-q_0q_2) &    2(q_0q_1+q_2q_3)        & q_0^2-q_1^2-q_2^2+q_3^2		
		\end{bmatrix}		
		\end{equation}	
		
		This can easily be seen as: $\vec{w}=2E\dot{q}=2E(G^T\vec{w}'/2) = EG^T\vec{w}'\implies R=EG^T$.
		
		The rotation matrix derivative can then easily be obtained as[1]
		\begin{eqnarray}
			\dot{R} &=& R\Omega'\\
			\Omega' &=& 2G\dot{G}^T
		\end{eqnarray}
		where $\Omega'$ is the skew symmetric matrix corresponding to the angular velocity vector.

	\section{Problems with Quaternions and other attitude representations}
	
	\subsection{Quaternions}
	\begin{equation*}
	(q_0,\vec{q}) = (-q_0,-\vec{q})
	\end{equation*}
	
	Since there are two quaternions:
		\begin{itemize}
			\item Ambiguities in representing an attitude: The constraint of unit modulus of quaternions restrict the quaternions to a sphere of unit radius in 4 dimensions, known as the three-sphere which is the set of unit-vectors in $R^4$. This three sphere doublecovers the attitude configuration of the special orthogonal group, SO(3).
			\item Single physical attitude of a rigid body may yield two different control inputs, which causes inconsistency in the resulting control system. A specific choice between two quaternions generates discontinuity that makes the resulting control system sensitive to noise and disturbances.
		\end{itemize}
		
		\subsection{General Representation}
	
	We first need to understand an important fact related to attitude stabilization. It has been proved mathematically[4] that there cannot exist a smooth controller on the configuration space (such as $SO(3)$ or $S^3$) that is globally asymptotically stable. 
	
	\textbf{Theorem}: The geometry of attitude space precludes the existence of continuous time invariant control that achieves global asymptotic stability.
	
	The proof of this theorem is a little involved with consideration of topological obstructions and configuration manifold which is out of scope of these notes. Details can be found in Reference [4]. Note that even the geometric controllers which work on $SO(3)$ directly instead of any parameterizations have also been shown to be almost globally asymptotically stable, such as in [5], where a two dimensional manifold is excluded from the three dimensional $SO(3)$ in the region of attraction. In case of quaternions or Euler angles it is easy to note that since $q$ and $-q$ and $0$ and $2\pi$ represent the same attitude, there exists more than one equilibrium point and thus one cannot even talk about global stability. One can filter the algorithm for these cases but then the controller becomes discontinuous (non-smooth) and hence the above theorem is not violated.
	
		\textbf{Unwinding Phenomenon}: The body may rest arbitrarily close to the desired attitude yet rotate through large angles before coming to rest in the desired attitude[4].
		
		Consider a body rotating about a single axis in a plane with angular velocity $\omega(t)$ with angular position $\theta(t)$. We can design a control torque $u(t)$ to stabilize the system dynamics as:
		\begin{eqnarray}
			\dot{\theta}(t) = \omega(t)\\
			\dot{\omega}(t) = u(t)\\
			u(t) = -k\theta(t) - c\omega(t)\\
			\implies \ddot{\theta}(t) + c\dot{\theta}(t) + k\theta(t) = 0
		\end{eqnarray}
		
		It may seem that $\theta=0$ is globally asymptotically stable point but values of $\theta$ that differ by integral values of $2\pi$ represent the same point in configuration space and hence it is not well defined in the state space. If we start at $\theta=2\pi-\epsilon$, then instead of going shorter way of $\epsilon$ anticlockwise, the controller will take it $2\pi-\epsilon$ clockwise.

	\section{Quaternion Error Dynamics}
	
	The quaternion error is defined as the relative rotation between the desired rotation and actual rotation. In order to better understand the inverse multiplication operation, we take example of rotation matrices. Suppose the rotation matrices corresponding to two consecutive relative rotations are $R_1$ and $R_2$. The net rotation matrix is then given by $R_3=R_2R_1$. Now if $R_3$ and $R_2$ were to be given, then $R_1$ can be found by multiplying both sides by inverse of $R_1$.
	\begin{eqnarray}
		R_3&=&R_2R_1\\
		\implies R_1 &=& R_2^{-1}R_3
	\end{eqnarray}
	Now suppose $R_3=R$ takes us from body to global frame and $R_2=R_d$ takes us from desired to global frame. Then $R_1=\tilde{R}$(error rotation matrix) takes us from body to desired frame. Hence the error rotation matrix can be found out by:
	\begin{equation}
		\tilde{R} = R_d^{-1}R
	\end{equation}
	The quaternions behave in a similar way to the rotation matrices. If the body attitude quaternion is $q$ and the desired body attitude is $q_d$ then the the error quaternion is defined as
	\begin{equation}
		q_e = q_d^{-1}\circ q
	\end{equation}
	As verification, note that
	\begin{eqnarray}
		q &=& q_d\circ q_e  \label{qedef} \\
		x &=& q \circ x' \circ q^*\\
		 &=& q_d \circ q_e \circ x' \circ q_e^*\circ q_d^*\\
		 &=& q_d \circ x_d' \circ q_d^*\\ ~&&(since ~ error ~quaternion~ takes~ a ~vector~ from~ body~ frame~ to~ desired ~frame)\\
		 &=& x \\~&& (since ~q_d~ takes~ a ~vector~ from ~desired~ frame ~to ~world ~frame)
	\end{eqnarray}
	This is just one way to define the error quaternion and authors may define this quaternion in different ways. 
	
	\subsection{Quaternion Error dynamics: First Order}
	
	Closed loop systems compute a feedback based not only on the error of involved quantities but also many times based on the rate of change of error. For this purpose, we derive here the expressions for quaternion error derivatives. It is to be noted that subtraction between two quantities has a meaning(or is consistent) only if they are expressed in same frame. Hence, suppose we are given a history of desired attitudes and we obtain the angular velocity from it. This operation will return us the derivative in the desired frame of references at each instant. To compare it to actual angular velocity, defined in the body frame, appropriate transformations have to be made which will be become clear below. Note that these transformations arise naturally since the physical quantities have to be consistent in their arithmetic.
	
	From Eq.(\ref{qedef}), we can obtain the time derivative of the error quaternion in the following way
		\begin{eqnarray}
		q_e &= &q_d^{-1}\circ q\\
		q_d\circ q_e &=& q\\
		\dot{q}_d\circ q_e + q_d\dot{q}_e &=& \dot{q}\\
		\implies \dot{q}_e &=& \bar{q}_d\circ (\dot{q}-\dot{q}_d\circ q_e)\\
		&=& \bar{q}_d\circ ( \frac{1}{2}q\circ w^B - \frac{1}{2}q_d\circ w_d^D\circ q_e )\\
	&&	(using~ derivative~ of~ quaternion~ expressionEq.(\ref{qder})\\
	&& ~for~ both~ body ~frame ~and ~desired~ frame ~quaternion)\\
		&=& \frac{1}{2} (\bar{q}_d\circ q\circ w_B - w_d^D\circ q_e)\\
		&=& \frac{1}{2}(q_e \circ w^B - w_d^D\circ q_e)\\
		&=& \frac{1}{2}q_e\circ (w^B - \bar{q}_e\circ w_d^D\circ q_e)
		\end{eqnarray}
		
		where $w_d$ is the desired angular velocity quaternion and $D$ stands for desired quaternion frame and $B$ denotes body frame.		
		Now, according to $  x = q\circ x'\circ \bar{q} $, in  $ q = q_d\circ q_e $ \\
	\ \  \ \  \ \  \  \ \  \ 	q: B to G frame, \ \ \ \ \ \ 	$ q_d $: D to G frame \ \ \ \ \ \  $ 	\implies q_e $: B to D frame
		\begin{equation}
	\therefore w_d^D = q_e\circ w_d^B \circ \bar{q}_e 
		\implies \dot{q}_e = \frac{1}{2}q_e\circ (w^B - w_d^B)
		\end{equation}		
		
		Denoting the angular velocity error as $\tilde{w}=w^B-w_d^B$, the above equation can be written in a similar form as Eq.(\ref{qder})
		\begin{equation}
				\dot{q}_e = \frac{1}{2}q_e\circ \tilde{w}
				\label{qeder2}
		\end{equation}

 \section{Quaternion to Euler Angle Conversion}
 
 Euler angles can be defined in many ways depending on the order of rotation and are still in popular use due to their simple physical interpretations. However it can be tricky to convert between quaternions and the Euler angles and this section gives a method to derive this conversion. The formula derived can be verified in a number of resources[7].
 
 Let XYZ represent the rotation first about X axis by $\phi$, then about new Y axis by $\theta$ and finally about new Z axis by $\psi$. Then the rotation matrix converting a vector from final frame (x,y,z) to original frame(X,Y,Z) is given by
 \begin{eqnarray}
 	\begin{bmatrix}
 	X\\Y\\Z
 	\end{bmatrix} &=& R_{\phi}R_{\theta}R_{\psi}\begin{bmatrix}
 	x\\y\\z
 	\end{bmatrix}\\
 	 &=& \begin{bmatrix}
 	   1 & 0 & 0\\
 	   0 & \cos\phi & -\sin\phi\\
 	   0 & \sin\phi & \cos\phi 
 	\end{bmatrix}\begin{bmatrix}
 	\cos\theta & 0 & \sin\theta\\
 	0 & 1 & 0\\
 	-\sin\theta & 0 & \cos\theta
 	\end{bmatrix}\begin{bmatrix}
 	\cos\psi & -\sin\psi & 0\\
 	\sin\psi & \cos\psi & 0\\
 	0 & 0 & 1 
 	\end{bmatrix}\begin{bmatrix}
 	x\\y\\z
 	\end{bmatrix}\\
 	&=& \begin{bmatrix}
 	    \cos\theta\cos\psi & -\cos\theta\sin\psi & \sin\theta \\
 	    \cos\phi\sin\psi+\sin\phi\sin\theta\cos\psi & \cos\phi\cos\psi-\sin\phi\sin\theta\sin\psi & -\sin\phi\cos\theta \\
 	    \sin\phi\sin\psi-\cos\phi\sin\theta\cos\psi & \sin\phi\cos\psi+\cos\phi\sin\theta\sin\psi & \cos\phi\cos\theta
 	\end{bmatrix}\begin{bmatrix}
 	x\\y\\z
 	\end{bmatrix}
 \end{eqnarray}
 
 We have already derived the rotation matrix with quaternions which is
 
 	\begin{equation}
 	R = \begin{bmatrix}
 	q_0^2+q_1^2-q_2^2-q_3^2 &   2(q_1q_2-q_0q_3)   &      2(q_0q_2+q_1q_3)\\
 	2*(q_1q_2+q_0q_3)&   (q_0^2-q_1^2+q_2^2-q_3^2) &   2(q_2q_3-q_0q_1)\\
 	2(q_1q_3-q_0q_2) &    2(q_0q_1+q_2q_3)        & q_0^2-q_1^2-q_2^2+q_3^2		
 	\end{bmatrix}		
 	\end{equation}
 	
 	Comparing the above two matrices element by element, we can find Euler angles from the quaternions:
 	\begin{eqnarray}
 		\phi = \tan^{-1}\left(\frac{-2(q_2q_3-q_0q_1)}{q_0^2-q_1^2-q_2^2+q_3^2}\right)\\
 		\theta = \sin^{-1}\left(2(q_0q_2+q_1q_3)\right)\\
 		\psi = \tan^{-1}\left(\frac{-2(q_1q_2-q_0q_3)}{q_0^2+q_1^2-q_2^2-q_3^2}\right)
 	\end{eqnarray}
 	
 	To obtain quaternion in terms of Euler Angles, we can solve the above system of equations. However, it can be tedious to solve. Instead, we can directly construct quaternion from the knowledge of Euler sequence. For the above sequence, just like we proceeded while deriving the quaternion derivative, the local rotation is compounded to the right. We first represent the quaternions for three rotations as
 	\begin{equation}
 	   q_{\phi} = \begin{bmatrix}
 	   \cos\left(\frac{\phi}{2}\right)\\
 	   \sin\left(\frac{\phi}{2}\right)\\
 	   0\\
 	   0
 	   \end{bmatrix}, ~~ q_{\theta} = \begin{bmatrix}
 	   \cos\left(\frac{\theta}{2}\right)\\
 	   0\\
 	   \sin\left(\frac{\theta}{2}\right)\\
 	   0
 	   \end{bmatrix}, ~~ q_{\psi} = \begin{bmatrix}
 	   \cos\left(\frac{\psi}{2}\right)\\
 	   0\\
 	   0\\
 	   \sin\left(\frac{\psi}{2}\right)
 	   \end{bmatrix}
 	\end{equation}
 	
 	Now compounding the rotations as before, i.e., post-multiplying the local rotations, we get
 	\begin{eqnarray}
 		q &=& q_{\phi}\circ q_{\theta}\circ q_{\psi}\\
 		 &=& \begin{bmatrix}
 		 \cos(\phi/2)\cos(\theta/2)\cos(\psi/2)-\sin(\phi/2)\sin(\theta/2)\sin(\psi/2)\\
 		 \cos(\phi/2)\sin(\theta/2)\sin(\psi/2)+\sin(\phi/2)\cos(\theta/2)\cos(\psi/2)\\
 		 \cos(\phi/2)\cos(\psi/2)\sin(\theta/2)-\sin(\phi/2)\cos(\theta/2)\sin(\psi/2)\\
 		 \cos(\phi/2)\cos(\theta/2)\sin(\psi/2)+\cos(\psi/2)\sin(\theta/2)\sin(\phi/2)
 		 \end{bmatrix}
 	\end{eqnarray}
 	The above expressions can be verified with the NASA documentation[7]. The reader can also refer to this document for transformations of all 12 Euler angle combinations.

 \section*{Disclaimer}
 These notes on quaternions are by no means complete and exhaustive. They only describe some fundamental details and do not go into much depth. They serve only to introduce the quaternions and coherently state their properties so that the reader can now readily refer to other literatures and understand the subtlety involved in the use of quaternions due to the existence of multiple conventions. Interested readers may refer to the references given below and to the vast literature accessible on Internet for greater detail.
 
 \section*{Refernences}
 
 \ \ \ \ [1]Graf, Basile. "Quaternions and dynamics." arXiv preprint arXiv:0811.2889 (2008)
 
 [2]Sola, Joan. "Quaternion kinematics for the error-state KF." Laboratoire d’Analyse et d’Architecture des Systemes-Centre national de la recherche scientifique (LAAS-CNRS), Toulouse, France, Tech. Rep (2012)
 
 [3]Diebel, James. "Representing attitude: Euler angles, unit quaternions, and rotation vectors." Matrix 58.15-16 (2006): 1-35
 
 [4]Bhat, Sanjay P., and Dennis S. Bernstein. "A topological obstruction to continuous global stabilization of rotational motion and the unwinding phenomenon." Systems \& Control Letters 39.1 (2000): 63-70
 
 [5]Lee, Taeyoung, Melvin Leok, and N. Harris McClamroch. "Control of complex maneuvers for a quadrotor UAV using geometric methods on SE (3)." arXiv preprint arXiv:1003.2005 (2010)
 
 [6]Choset, Howie M. Principles of robot motion: theory, algorithms, and implementation. MIT press, 2005.
 
 [7]Henderson, D. M. "Euler angles, quaternions, and transformation matrices—Working relationships." NASA TM-74839, JSC-12960 (1977).

\end{document}